\documentclass[a4paper,11pt]{article}
\pdfoutput=1 

\usepackage{jcappub} 

\usepackage[T1]{fontenc} 

\usepackage{float}
\usepackage{hyperref}
\usepackage{tablefootnote}
\usepackage{footnotehyper}
\usepackage{graphicx}
\hypersetup{colorlinks, linkcolor = {blue} }
\usepackage{adjustbox}
\usepackage{amsmath, bm}

\usepackage[utf8]{inputenc}

\usepackage{CJK}

\usepackage[usenames,dvipsnames]{color} 

\usepackage{hologo} 
\usepackage{booktabs}

\graphicspath{{./}{figures/}}

\newcommand\aap{Astron. Astrophys.}

\newcommand\apjl{Astrophys. J. L.}
\newcommand\apjs{Astrophys. J. S.}

\title{Cosmological joint analysis with cosmic growth and expansion rate}

\author[a,b]{Jing~Niu}
\author[a,b,c]{Tong-Jie~Zhang}

\affiliation[a]{Institute for Frontiers in Astronomy and Astrophysics, Beijing Normal University, Beijing 102206, China}
\affiliation[b]{Department of Astronomy, Beijing Normal University, Beijing 100875, China}
\affiliation[c]{Institute for Astronomical Science, Dezhou University, Dezhou 253023, China}

\emailAdd{tjzhang@bnu.edu.cn}

\abstract{The measurements of expansion rate $H(z)$ and the growth rate $f\sigma_8(z)$ describe the evolution of the universe, and both of them can constrain the cosmological models through data analysis. Due to the lack of data points, these datasets are combined by the traditional combined method ($\chi^2$ method) to select a best-fitting cosmological model. In 2017, Linder proposed a joint method, which describes the evolution of the universe through $H(z)-f\sigma_8$ diagram instead of the redshift z. Compared to individual datasets, Linder demonstrated the advantages of the joint method to distinguish cosmologies. In this paper, we compare the significance between the traditional combined method and Linder's joint method by constraining the density parameter $\Omega_M$ using Akaike Information Criterion (AIC) and Bayesian Information Criterion (BIC). The result shows that the joint method is more significant than the traditional combined method.}

\begin{document}
\maketitle
\flushbottom

\section{Introduction}
\label{sec:intro}

Constraining parameters to select a best-fitting cosmological model that matches observed datasets is critical in cosmology.  
There are lots of independent cosmological probes, such as the supernova (SN) observations \citep[see, e.g.,][]{2009ApJ...700.1097H,2018Natur.554..497B}, baryon acoustic oscillations (BAO) \citep[see, e.g.,][]{2020ApJ...904L..17P}, cosmic microwave background (CMB) \citep[see, e.g.,][]{2021ApJ...912L...1A}, integrated Sachs-Wolfe effect \citep[see, e.g.,][]{2008ApJ...683L..99G}, galaxy clusters \citep[see, e.g.,][]{2017ApJ...846L..31D}, and strong gravitational lensing \citep[see, e.g.,][]{2017Natur.548..555H}. With all the probes, astronomers dedicate themselves to collecting more accurate date sets from the universe to discriminate the cosmological frameworks. Due to the lack of data, convincing conclusions can be drawn in combination with currently available datasets.

Linder proposed a joint method to analyze the datasets in 2017~\citep{paper:Linder2017}. They interpreted the data in a different form, the Hubble expansion rate $H(z)$ versus growth rate $f\mathrm{\sigma_8}$, to describe the evolutionary tracks. $H(z)=\dot{a}/a$ describes the expansion history, where $a=1/(1+z)$ is the cosmic scale factor. $f\sigma_8$ describes the growth history of the universe. $f\sigma_8$ is the product of the growth rate f and the root mean square of matter fluctuations $\sigma_8$, where $f=\mathrm{d} lnD/\mathrm{d} lna$, D is the linear growth factor. The variance of matter fluctuations $\mathrm{\sigma_8^2}=\frac{1}{2\pi^2}\int^\infty_0\mathrm{P_m}(\mathit{k,z})\mathrm{\widehat{W}^2}(\mathit{k})\mathit{k}^2\mathrm{d}\mathit{k}$, where $\mathrm{P_m}$ is the matter power spectrum, and $\mathrm{\widehat{W}}$ is a top-hat spherical window function of radius 8$h^{-1}$Mpc~\citep{paper:Edwin2005}.

Several papers used Linder's joint method to study cosmology from different perspectives. Such as ~\citet{paper:M.Moresco2017} constrained the possible extensions to the standard flat $\Lambda$CDM model, and they got the best-fitting growth index was $\gamma=0.65^{+0.05}_{-0.04}$ and the best-fitting total neutrino mass was $\sum m_{\nu}=0.26\pm0.10$ev. \citet{paper:SpyrosBasilakos2017} constrained several cosmological models with Linder's joint method. And they introduced the Figure of Merit (FoM), which is the inverse of the enclosed area of the 1 $\sigma$ errors of the best-fitting models with datasets. FoM could quantify the power of constraining, the FoM higher, the more constraining. According to this, they found that the $\Lambda$CDM model is the most constraining model. All of the above studies just used the joint method to choose a universal model, they didn’t consider the significance of the joint method. The works we mentioned above without the significance of the joint method does not show the relative quality of the joint method relative to other methods commonly used in the field of astronomy. It is useless to develop a joint method if the quality of the joint method is not as good as that of the traditional method. In this paper, we compare the significance of joint method with traditional combined method through AIC and BIC. 


This work is organized as follows: In section \ref{sec:data} we compile the available independent datasets of $H(z)$ and $f\sigma_8$. Section \ref{sec:methods} introduces the traditional combined method ($\chi^2$ method) and joint method, and constrain $\mathrm{\Omega_M}$ using both methods. We compare the significance of the two methods using AIC and BIC in section \ref{sec:analysis}. Finally, section \ref{sec:conclusion} is the conclusion and discussion of this paper.

\section{Data}
\label{sec:data}

We constrain the expansion universe using the dataset from the cosmic chronometers (CC) approach. In the CC method, the expansion rate is calculated by the differential ages of galaxies, given by
\begin{equation}
H(z)=-\frac{1}{1+z}\frac{\mathrm{d}z}{\mathrm{d}t}.
\label{eq:1}
\end{equation}
High-precision $\mathrm{d}z$ can be obtained from spectroscopic surveys and $\mathrm{d}t$ is the differential age in a given $\mathrm{d}z$, so CC is a cosmological model-independent technique compared to other standard probes (e.g. BAO, SNE).
There are several models to fit the observed spectra, such as BC03 \citep[see, e.g.,][]{paper:BruzualG.2003}, PE \citep[see, e.g.,][]{paper:LeBorgneD.2004}, VM\citep[see, e.g.,][]{paper:VazdekisA.2010}, M11\citep[see, e.g.,][]{paper:MarastonC.2011}, etc. These models provide slightly different $H(z)$ estimates. The $H(z)$ dataset considered in this work is from BC03 measurements, it is widely used in this field. All CC data we used in this work are compiled in Table \ref{tab:hz}.

\begin{table}
\begin{minipage}{\textwidth}
	\centering\caption{The set of available observational $H(z)$ data (CC)}
	
	\begin{tabular*}{.52\textwidth}{lccc}
		\hline\hline
		~~~Redshift $z$ &\hspace*{0em} $H(z)\footnote{$H(z)$ figures are in the unit of kms$^{-1}$ Mpc$^{-1}$}$$\pm 1\sigma$ error &\hspace*{0em}References \\
		
		\hline
		~~~~~~0.07 &\hspace*{0em} 69 $\pm$ 19.6  & \hspace*{0em}\cite{paper:C.Zhang2014} \\
		~~~~~~0.1 &\hspace*{0em} 69 $\pm$ 12  & \hspace*{0em}\cite{paper:JSimon2005} \\
        ~~~~~~0.12 &\hspace*{0em} 68.6 $\pm$ 26.2   & \hspace*{0em}\cite{paper:C.Zhang2014} \\
        ~~~~~~0.17 & \hspace*{0em}83 $\pm$ 8  & \hspace*{0em}\cite{paper:JSimon2005}\\
        ~~~~~~0.1791 & \hspace*{0em}75 $\pm$ 5  & \hspace*{0em}\cite{paper:MorescoM.2012} \\
        ~~~~~~0.1993 & \hspace*{0em}75 $\pm$ 5  & \hspace*{0em}\cite{paper:MorescoM.2012}\\
        ~~~~~~0.2  &\hspace*{0em}72.9 $\pm$ 29.6  & \hspace*{0em}\cite{paper:C.Zhang2014}\\
        ~~~~~~0.27 &\hspace*{0em} 77 $\pm$ 14  &\hspace*{0em}\cite{paper:JSimon2005}\\
        ~~~~~~0.28 &\hspace*{0em} 88.8 $\pm$ 36.6  & \hspace*{0em}\cite{paper:C.Zhang2014}\\
        ~~~~~~0.3519 &\hspace*{0em}83 $\pm$ 14  & \hspace*{0em}\cite{paper:MorescoM.2012}\\
        ~~~~~~0.3802 &\hspace*{0em}83 $\pm$ 13.5  & \hspace*{0em}\cite{paper:MorescoM.2016a}\\
        ~~~~~~0.4  &\hspace*{0em}95 $\pm$ 17 & \hspace*{0em}\cite{paper:JSimon2005}\\
        ~~~~~~0.4004 &\hspace*{0em}77 $\pm$ 10.2  & \hspace*{0em}\cite{paper:MorescoM.2016a}\\
        ~~~~~~0.4247  &\hspace*{0em}87.1 $\pm$ 11.2 & \hspace*{0em}\cite{paper:MorescoM.2016a}\\
        ~~~~~~0.4497 &\hspace*{0em}92.8 $\pm$ 12.9 & \hspace*{0em}\cite{paper:MorescoM.2016a}\\
        ~~~~~~0.47 &\hspace*{0em}89 $\pm$ 67  & \hspace*{0em}\cite{paper:A.L.Ratsimbazafy2017}\\
        ~~~~~~0.4783 &\hspace*{0em}80.9 $\pm$ 9 & \hspace*{0em}\cite{paper:MorescoM.2016a}\\
        ~~~~~~0.48 &\hspace*{0em} 97 $\pm$ 62 & \hspace*{0em}\cite{paper:D.Stern2010}\\
        ~~~~~~0.5929 &\hspace*{0em} 104 $\pm$ 13 & \hspace*{0em}\cite{paper:MorescoM.2012}\\
        ~~~~~~0.6797 &\hspace*{0em} 92 $\pm$ 8 & \hspace*{0em}\cite{paper:MorescoM.2012}\\
        ~~~~~~0.7812 &\hspace*{0em} 105 $\pm$ 12 & \hspace*{0em}\cite{paper:MorescoM.2012}\\
        ~~~~~~0.8754 &\hspace*{0em} 125 $\pm$ 17 & \hspace*{0em}\cite{paper:MorescoM.2012}\\
        ~~~~~~0.88 &\hspace*{0em} 90 $\pm$ 40 & \hspace*{0em}\cite{paper:D.Stern2010}\\
        ~~~~~~0.9 &\hspace*{0em} 117 $\pm$ 23 & \hspace*{0em}\cite{paper:JSimon2005}\\
        ~~~~~~1.037 &\hspace*{0em} 154 $\pm$ 20 & \hspace*{0em}\cite{paper:MorescoM.2012}\\
        ~~~~~~1.3 &\hspace*{0em} 168 $\pm$ 17 & \hspace*{0em}\cite{paper:JSimon2005}\\
        ~~~~~~1.363 &\hspace*{0em} 160 $\pm$ 33.6 & \hspace*{0em}\cite{paper:MorescoM.2015}\\
        ~~~~~~1.43 &\hspace*{0em} 177 $\pm$ 18 & \hspace*{0em}\cite{paper:JSimon2005}\\
        ~~~~~~1.53 &\hspace*{0em} 140 $\pm$ 14 & \hspace*{0em}\cite{paper:JSimon2005}\\
        ~~~~~~1.75 &\hspace*{0em} 202 $\pm$ 40 & \hspace*{0em}\cite{paper:JSimon2005}\\
        ~~~~~~1.965 &\hspace*{0em} 186.5 $\pm$ 50.4 & \hspace*{0em}\cite{paper:MorescoM.2015}\\
		\hline\hline
	\end{tabular*}
    \label{tab:hz}
\end{minipage}
\end{table}


Friedmann equation describes the expansion of the universe.
\begin{equation}
        H(z)=\mathrm{H_0}\sqrt{\mathrm{\Omega_M}(1+z)^3+\mathrm{\Omega_R}(1+z)^4+\mathrm{\Omega_k}(1+z)^2+\Omega_\Lambda(1+z)^{3(1+w)}}
        \label{eq:2}
\end{equation}
In this work, we consider the $\mathrm{\Lambda CDM}$ cosmology, assuming the universe is flat $\mathrm{\Omega_k=0}$ and $w=1$. Since $\mathrm{\Omega_R\ll\Omega_M}$, we neglect $\mathrm{\Omega_R}$. The Friedmann equation for this flat $\mathrm{\Lambda CDM}$ cosmology is
\begin{equation}
        H(z)=\mathrm{H_0}\sqrt{\mathrm{\Omega_M}(1+z)^3
+(1-\mathrm{\Omega_M})}.
        \label{eq:3}
\end{equation}
Assuming in $\mathrm{\Lambda CDM}$ cosmology, \citet{paper:N.Aghanimetal2020} gave $H_{0}=67.4 \pm 0.5 $ kms$^{-1}$ Mpc$^{-1}$. We use equation (\ref{eq:3}) to constrain $\mathrm{\Omega_M}$.

Concerning the cosmic growth, a large amount of $f\sigma_8$ data are reported. But we can't use them all to constrain the cosmological parameters, only the independent $f\sigma_8$ is included in our work. The $f\sigma_8$ data points we used in our analysis are compiled in Table \ref{tab:fsigma8}. We analyze $f\sigma_8$ using the method proposed by \citet{paper:LavrentiosKazantzidis2018} and constrain $\mathrm{\Omega_M}$ using the $f\sigma_8$ dataset in $\mathrm{\Lambda}$CDM cosmology.

\begin{table}
	\centering\caption{The compiled independent $f\sigma_8$ dataset}
	\begin{tabular*}{.52\textwidth}{lccc}
		\hline\hline
		~~~Redshift $z$ &\hspace*{0em} $f\sigma_8(z)\pm 1\sigma$ error  &\hspace*{0em}References \\
		
		\hline
		~~~~~~0.02 &\hspace*{0em} 0.428 $\pm$ 0.0465  &\hspace*{0em} \cite
		{paper:DraganHuterer2016} \\
		~~~~~~0.02 &\hspace*{0em} 0.398 $\pm$ 0.065  &\hspace*{0em} \cite{paper:StephenJ.Turnbull2012} \\
        ~~~~~~0.02 &\hspace*{0em} 0.314 $\pm$ 0.048   & \hspace*{0em}\cite{paper:MichaelJ.Hudson2013} \\
        ~~~~~~0.10 & \hspace*{0em}0.370 $\pm$ 0.130  & \hspace*{0em}\cite{paper:MartinFeix2015}\\
        ~~~~~~0.15 & \hspace*{0em}0.490 $\pm$ 0.145  & \hspace*{0em}\cite{paper:CullanHowlett2015} \\
        ~~~~~~0.17 & \hspace*{0em}0.510 $\pm$ 0.060  & \hspace*{0em}\cite{paper:Yong-SeonSong2009}\\
        ~~~~~~0.18  &\hspace*{0em}0.360 $\pm$ 0.090  & \hspace*{0em}\cite{paper:ChrisBlakeetal2013}\\
        ~~~~~~0.25 &\hspace*{0em} 0.3512 $\pm$ 0.0583  &\hspace*{0em}\cite{paper:LadoSamushia2012}\\
        ~~~~~~0.32 &\hspace*{0em} 0.384 $\pm$ 0.095  & \hspace*{0em}\cite{paper:ArielG.Sanchezetal2014}\\
        ~~~~~~0.37 &\hspace*{0em}0.4602 $\pm$ 0.0378  & \hspace*{0em}\cite{paper:LadoSamushia2012}\\
        ~~~~~~0.38  &\hspace*{0em}0.440 $\pm$ 0.060 & \hspace*{0em}\cite{paper:ChrisBlakeetal2013}\\
        ~~~~~~0.44 &\hspace*{0em}0.413 $\pm$ 0.080  & \hspace*{0em}\cite{paper:ChrisBlakeetal2012}\\
        ~~~~~~0.59 &\hspace*{0em}0.488 $\pm$ 0.060  & \hspace*{0em}\cite{paper:Chia-HsunChuangetal2016}\\
        ~~~~~~0.60  &\hspace*{0em}0.390 $\pm$ 0.063 & \hspace*{0em}\cite{paper:ChrisBlakeetal2012}\\
        ~~~~~~0.60 &\hspace*{0em}0.550 $\pm$ 0.120 & \hspace*{0em}\cite{paper:A.Pezzottaetal2016}\\
        ~~~~~~0.72 &\hspace*{0em}0.454 $\pm$ 0.139  & \hspace*{0em}\cite{paper:MIcaza-Lizaolaetal2020}\\
        ~~~~~~0.73 &\hspace*{0em}0.437 $\pm$ 0.072 & \hspace*{0em}\cite{paper:ChrisBlakeetal2012}\\
        ~~~~~~0.86 &\hspace*{0em} 0.400 $\pm$ 0.110 & \hspace*{0em}\cite{paper:A.Pezzottaetal2016}\\
        ~~~~~~0.978 &\hspace*{0em} 0.379 $\pm$ 0.176 & \hspace*{0em}\cite{paper:Gong-BoZhaoetal2018}\\
        ~~~~~~1.23 &\hspace*{0em} 0.385 $\pm$ 0.099 & \hspace*{0em}\cite{paper:Gong-BoZhaoetal2018}\\
        ~~~~~~1.40 &\hspace*{0em} 0.482 $\pm$ 0.116 & \hspace*{0em}\cite{paper:TeppeiOkumuraetal2016}\\
        ~~~~~~1.52 &\hspace*{0em} 0.420 $\pm$ 0.076 & \hspace*{0em}\cite{paper:HectorGil-Marinetal2018}\\
        ~~~~~~1.52 &\hspace*{0em} 0.396 $\pm$ 0.079 & \hspace*{0em}\cite{paper:JiaminHouetal2018}\\
        ~~~~~~1.526 &\hspace*{0em} 0.342 $\pm$ 0.070 & \hspace*{0em}\cite{paper:Gong-BoZhaoetal2018}\\
        ~~~~~~1.944 &\hspace*{0em} 0.364 $\pm$ 0.106 & \hspace*{0em}\cite{paper:Gong-BoZhaoetal2018}\\
		\hline\hline
	\end{tabular*}
	\label{tab:fsigma8}
\end{table}

\begin{figure*}
\centering
\setlength{\floatsep}{8 pt plus 2 pt minus 3 pt}
\includegraphics[width=15cm,height=7.5cm]{{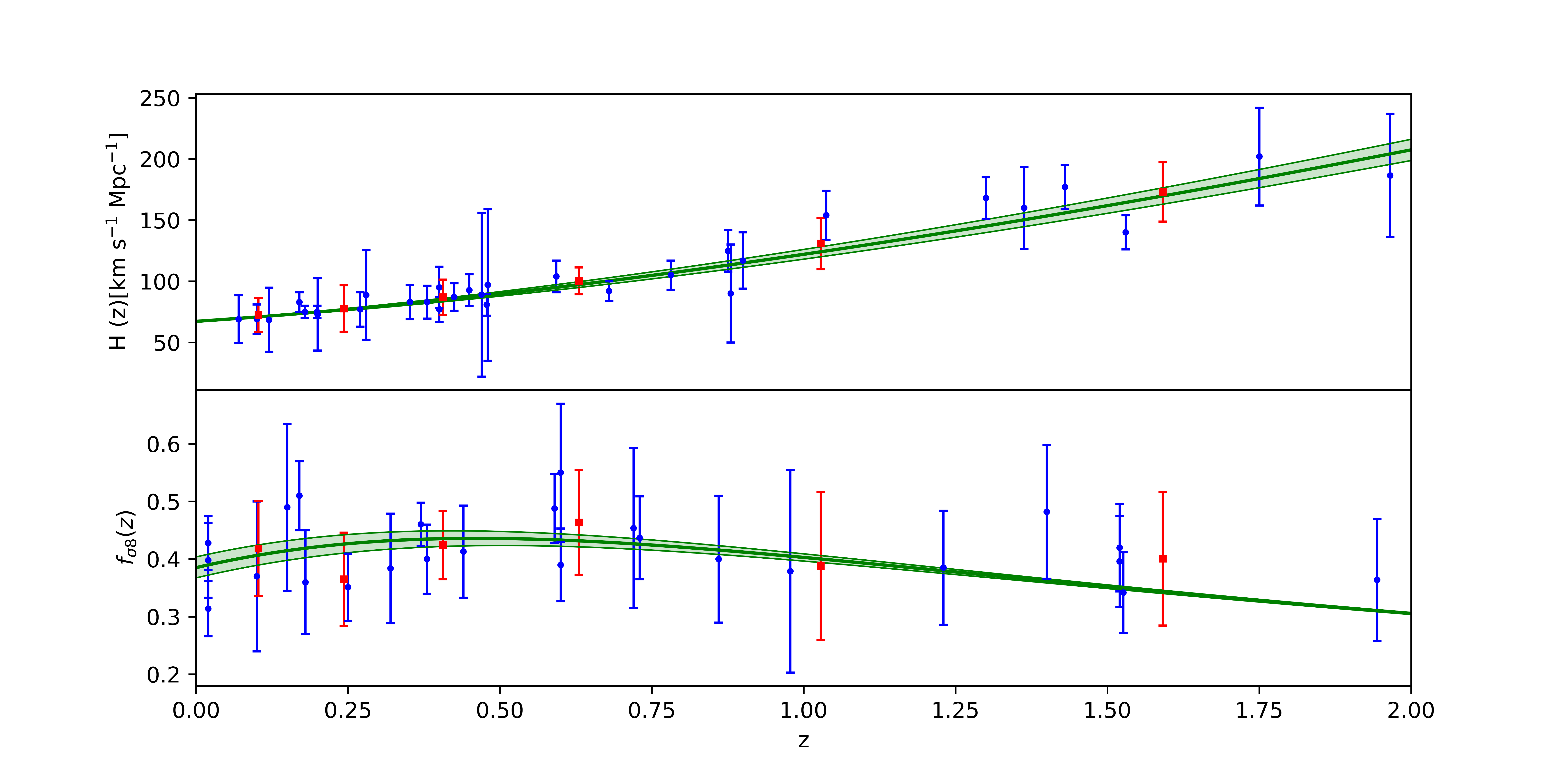}}
\caption{\label{f:1}The upper and lower panel shows the Hubble expansion parameter $H(z)$ and the growth rate $f\sigma_8$ respectively. The blue circular points show the data with $1\sigma$ associated errors we used in our work, which we compiled in Table\ref{tab:hz} and Table\ref{tab:fsigma8}. The red square points show the binned data with $1\sigma$ associated errors taht we analyzed in section \ref{subsec:jointmethod}. The green line and shaded area show the best-fitting $\mathrm{\Omega_M}$ and $1\sigma$ associated errors in $\mathrm{\Lambda CDM}$ model respectively. According to the dataset that we compiled in Table\ref{tab:hz} and Table\ref{tab:fsigma8}, the best-fitting matter density for expansion rate $H(z)$ is $\mathrm{\Omega_M}=0.333^{+0.031}_{-0.030}$, for growth rate $f\sigma_8$ is  $\mathrm{\Omega_M}=0.289^{+0.018}_{-0.017}$.} 
\end{figure*}

\section{Methods}
\label{sec:methods}

\subsection{traditional combined method ($\chi^2$ method)}
\label{subsec:traditional combined method}

The traditional combined method to constrain the cosmological parameters is $\chi^2$ method. We compile the independent Hubble dataset $H_\mathrm{{obs}}(z)$ in Table \ref{tab:hz} as $\bm{H_{\mathrm{obs}}}=(H_{\mathrm{obs1}},\cdot,\cdot,\cdot,H_{\mathrm{obsN}})^T$ at given redshift $\bm{z}=(z_1,\cdot\cdot\cdot,z_\mathrm{N})^T$ with error $\bm{\sigma_{\mathrm{obs}}}=(\sigma_{\mathrm{obs1}},\cdot\cdot\cdot,\sigma_{\mathrm{obsN}})^\mathrm{T}$. We consider the flat $\mathrm{\Lambda}$CDM cosmology in our work, and the Hubble parameter is given by (\ref{eq:3}). To determine $\mathrm{\Omega_M}$, we need to maximize the likelihood function based on  Bayes' theorem:
\begin{equation}
        P(\mathrm{\Omega_M}|H_{\mathrm{obs}})=\frac{P(H_{\mathrm{obs}}|\Omega_\mathrm{M})P(\mathrm{\Omega_M})}{P(H_{\mathrm{obs}})}.
        \label{eq:4}
\end{equation}
The likelihood function $\mathcal{L}$ is
\begin{equation}
        \mathcal{L}(\bm{H_{\mathrm{obs}}}|\mathrm{\Omega_M})=\displaystyle\prod_{i=1}^{N}P(H_{\mathrm{obsi}}|\mathrm{\Omega_M}).
        \label{eq:5}
\end{equation}
Assuming that $\bm{\mathrm{\sigma_{obs}}}$ is Gaussian error distribution, and  they are independent of each other, the likelihood function is given by \cite{paper:Ma2011,paper:YuChenWang2021}:
\begin{equation}
        \mathcal{L}(\bm{H_\mathrm{{obs}}}|\mathrm{\Omega_M})=\left(\overset{N}{\underset{i=1}{\prod}}\frac{1}{\sqrt{2\pi\sigma_i^2}}\right)exp\left(-\frac{\chi^2}{2}\right),
        \label{eq:6}
\end{equation}
where the $\chi^2$ statistic is 
\begin{equation}
        \chi^2=\underset{i}\sum\frac{[H(z_i,\mathrm{\Omega_M})-H_{\mathrm{obs},i}]^2}{\sigma_i^2},
         \label{eq:7}
\end{equation}
the $\chi^2_{f_{\sigma8}}$ for $f\sigma8$ is similar to $\chi^2_\mathrm{H}$ and the total $\chi^2_{\mathrm{tot}}$ is
\begin{equation}
       \chi^2_{\mathrm{tot}}=\chi^2_\mathrm{H}+\chi^2_{f_{\sigma8}}.
         \label{eq:8}
\end{equation}
We obtain the best-fitting $\mathrm{\Omega_M}=0.300^{+0.022}_{-0.022}$ for the $\chi^2$ method in \href{https://emcee.readthedocs.io/en/stable/}{EMCEE} by maximizing the likelihood function:
\begin{equation}
       \frac{\mathrm{d}\mathrm{log}\mathcal{L}}{\mathrm{d}\mathrm{\Omega_M}}=0.
         \label{eq:9}
\end{equation}

\subsection{joint method}
\label{subsec:jointmethod}

It is difficult to accurately constrain cosmological parameters by a single dataset, such as $H(z)$ for expansion rate and $f\sigma_8(z)$ for growth rate. Figure \ref{f:1} shows that H(z) is difficult to discriminate cosmologies at low redshift, and $f\sigma_8(z)$ is difficult to discriminate cosmologies at high redshift. Furthermore, in Figure \ref{f:1}, H(z) and $f\sigma_8(z)$ smoothly, slowly change with increasing redshift, and they have similar shapes in different cosmology. So we need to find an effective and visual method to discriminate the different cosmologies with these datasets.

\begin{figure}
\begin{center}
\setlength{\floatsep}{8 pt plus 2 pt minus 3 pt}
\includegraphics[width=8.9cm,height=6cm]{{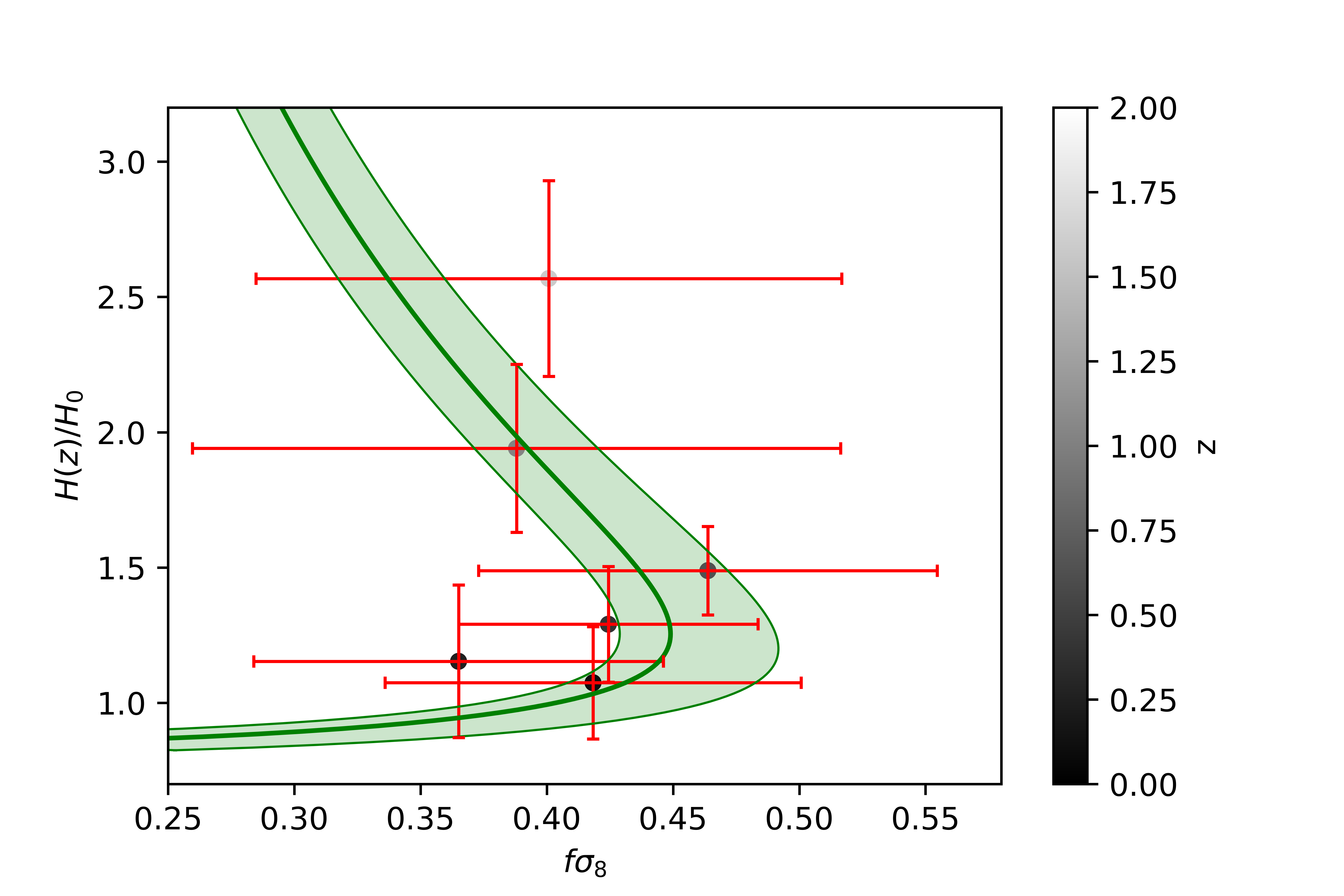}}
\caption{\label{f:2}The diagram of expansion rate $H(z)/\mathrm{H_0}$ against growth rate $f\sigma_8$. The circular points with $1\sigma$ associated errors are the bins which be calculated in section \ref{subsec:jointmethod}. The green line is the best-fitting $\mathrm{\Omega_M}=0.307^{+0.059}_{-0.053}$ using the bins in the $\mathrm{\Lambda CDM}$ model, and the shaded area is the $1\sigma$ associated errors. The color of the bins shows the redshift of each bin, the value of the redshift is shown in the color bar.}
\end{center}
\end{figure}

In 2017, Linder proposed the joint method to constrain the cosmological parameters, which has the advantage of distinguishing the different cosmologies \citep{paper:Linder2017}. Instead of $H(z)$ and $f\sigma_8(z)$ as a function of redshift, joint method describes the diagram of $H(z)$/$\mathrm{H_0}$ versus $f\sigma_8(z)$. Figure \ref{f:2} shows a wiggle in rage $z\in[0.4,1]$, cosmologies can be discriminated according to the degree and location of the wiggle. Thus, the joint method is effective to discriminate the cosmologies and constrain the universe.

To construct the $H(z)$/$\mathrm{H_0}$-$f\sigma_8(z)$, we collect the currently available datasets of $H(z)$ and $f\sigma_8(z)$  shown in Table \ref{tab:hz} and Table \ref{tab:fsigma8}, respectively. First of all, we need to bin the datasets of $H(z)$ and $f\sigma_8(z)$. There are different ways to bin them, such as equal intervals, where the range of each bin is the same \citep[see, e.g.,][]{paper:SpyrosBasilakos2017}. And unequal intervals, grouping the data points depends on the density of the data points. We count the total number of the data points of $H(z)$ and $f\sigma_8(z)$ as $N_\mathrm{{tot}}$. We set up 6 bins in this work, and 6 is the optimal number by experience.  And we calculate the number of each group, $N_\mathrm{{tot}}$/6. Then mix all the data of $H(z)$ and $f\sigma_8(z)$, then sort them, and group each data point of $H(z)$, $f\sigma_8(z)$ by the value of z. We calculate $z_\mathrm{{bin}}$ for each group of $H(z)$ and $f\sigma_8(z)$, respectively. To get the $H_\mathrm{{bin}}$ and $f\sigma_\mathrm{{8bin}}$ for each group, we compute the weighted average in Python
\begin{equation}
       H_\mathrm{{bin}}(z_\mathrm{{bin}})=\frac{\displaystyle\sum_{i\in \mathrm{bin}}\frac{H_i(z)}{z_i-z_\mathrm{{bin}}}}{\displaystyle\sum_{i\in \mathrm{bin}}\frac{1}{z_i-z_\mathrm{{bin}}}}.
         \label{eq:10}
\end{equation}
Similar to $f\sigma_\mathrm{{8bin}}(z_\mathrm{{bin}})$ and their errors. Finally, we constrain the $\mathrm{\Omega_M}=0.307^{+0.059}_{-0.053}$ with the binned dataset shown in Figure \ref{f:2}.

\section{Analysis}
\label{sec:analysis}

Figure \ref{f:1} compares the best-fitting $\mathrm{\Omega_M}$ constrained by the $H(z)$ and $f\sigma_8$ datasets, which are compiled in Table \ref{tab:hz} and Table \ref{tab:fsigma8}, respectively. The best-fitting cosmological parameter constrained by $H(z)$ and $f\sigma_8$ datasets in the traditional combined method is calculated in section \ref{subsec:traditional combined method},  $\mathrm{\Omega_M}=0.300^{+0.022}_{-0.022}$. We bin the data in Table \ref{tab:hz}, Table \ref{tab:fsigma8} and constrain $\mathrm{\Omega_M}=0.307^{+0.059}_{-0.053}$ with binned data using joint method in Figure \ref{f:2}. To compare the significance of the traditional combined method and joint method, we operate two kinds of selection model criteria, Akaike Information Criterion (hereafter AIC) \citep{paper:StoicaP.2004} and Bayesian information criterion (hereafter BIC) \citep{paper:Schwarz1978}.

Both AIC and BIC estimate the quality of the model with the given dataset, $\mathrm{\Delta AIC}$ and $\mathrm{\Delta BIC}$ give the relative quality between two moels. AIC and BIC estimate the missing information of a given model. The smaller the value of AIC and BIC, the less information the model loses and the higher the quality of the model. AIC and BIC consider both the goodness of fit and the simplicity of the  model. Both AIC and BIC suffer from overfitting, which they solve by adding a penalty term to the model. The difference is that the penalty term in BIC is larger than in AIC. The definitions of AIC and BIC are
\begin{equation}
       AIC=2k - 2\mathrm{ln}(\hat{L}),
         \label{eq:11}
\end{equation}
\begin{equation}
       BIC=k\mathrm{ln}(n) - 2\mathrm{ln}(\hat{L}).
         \label{eq:12}
\end{equation}
Where $\hat{L}$ is the maximum value of the likelihood function of the model, k is the number of the estimated parameters of the model, n is the sample size. Due to the small bin size, we modify the function for a small sample size. $\mathrm{AIC_c}$ \citep{paper:JosephE1997} is the correction for AIC with small sample size,

\begin{equation}
       AIC_c=AIC+\frac{2k^2+2k}{n-k-1}.
         \label{eq:13}
\end{equation}

$\mathrm{\Delta AIC_\mathrm{c}}$ and $\mathrm{\Delta BIC}$ between the traditional combined method and joint method are 
\begin{equation}
      \Delta AIC_\mathrm{c}=AIC_{\mathrm{c,combined}}-AIC_{\mathrm{c,joint}}=59.18,
            \label{eq:14}
\end{equation}
\begin{equation}
      \Delta BIC=BIC_{\mathrm{combined}}-BIC_{\mathrm{joint}}=66.11.
          \label{eq:15}
\end{equation}
The results of equations (\ref{eq:13}) and (\ref{eq:14}) show that the joint method is more significant than the traditional combined method.

\section{conclusions and discussion}
\label{sec:conclusion}



In this paper, we collect the available datasets of $H(z)$ and $f\sigma_8$. We select independent datasets from them to constrain $\Omega_M$, which are compiled in Table \ref{tab:hz} and Table \ref{tab:fsigma8}. We introduce two methods in section \ref{sec:methods}, the traditional combined method and the joint method. Within the joint method, we introduce different ways to bin, unequal intervals bin is used in this work. The joint method is a new method to analyze the cosmological parameters, and so far, the previous works have not compared the significance with the traditional combined method. Without the relative quality of the joint method to the traditional combined method, it's meaningless to use the joint method rather than the traditional combined method to select the best-fitting universal model. So it's critical to analyze the significance of both methods. We constrain the matter density $\mathrm{\Omega_M}$ by both the traditional combined method and the joint method. To compare the significance of these methods, we utilize the selection model criteria $\mathrm{AIC_c}$ and BIC. The differences of $\mathrm{AIC_c}$ and BIC between two methods are $\Delta \mathrm{AIC_c}=59.18$ and $\Delta \mathrm{BIC}=66.11$, which indicates that the joint method is higher significant. Therefore, the joint method is superior than the combined method in constraining cosmological parameter $\mathrm{\Omega_M}$. 





In the future, as more data points are released, the accuracy of constraints will be higher, the accuracy of the significance of the traditional combined method and joint method will also be higher. Admittedly, only AIC and BIC are not enough to justify the joint method is better. we need other ways to determine the advantages or weaknesses of the joint method. In the future, we could choose other methods to calculate the significance to compare the joint method with traditional method, such as statistical significance to calculate the significance level $\alpha$ and p-value. And due to the lack of theoretical support for the joint method, we could navigate and extend the joint method on more different kinds of datasets theoretically in the future.


\acknowledgments
We thank  Kang Jiao, Yu-Jie Lian and Chang-Zhi Lu for useful discussions. This work was supported by the National Science Foundation of China (Grants No. 11929301) and National Key R$\&$D Program of China (2017YFA0402600).




\bibliographystyle{aasjournal}
\bibliography{example}

\end{document}